\documentstyle[epsfig, emulateapj5]{aastex}
\received{}
\revised{}
\accepted{}

\shorttitle{Statistics of the IGM}
\shortauthors{Fang \& White}

\reversemarginpar

\begin{document}
\title{Probing the statistics of the temperature-density relation of the IGM}
\author{Taotao Fang\altaffilmark{1}, Martin White}
\affil{Department of Astronomy, University of California, Berkeley, CA 94720}

\begin{abstract}
Gravitational instability induces a simple correlation between the
large and small scale fluctuations of the Ly-$\alpha$ flux spectrum.
However, non-gravitational processes involved in structure formation
and evolution will alter such a correlation.  In this paper we explore
how scatter in the temperature-density relation of the IGM reduces the
gravitationally induced scale-scale correlation. By examining whether
or not observations of the correlation are close to that predicted by
pure gravity, this puts constraints on the scatter in the $\rho-T$
relation and in turn on any physical process which would lead to
scatter, e.g.~strong fluctuations in the UV background or radiative
transfer effects.  By applying this method to high resolution Keck
spectra of Q~1422+231 and HS~1946+7658, we find the predicted
correlation signal induced by gravity, and the diminishing of this
correlation signal at small scales. This suggests extra physics
affects the small-scale structure of the forest, and we can constrain
the scatter in the $\rho-T$ relation to a conservative 20\% upper
limit. A crude model suggests, if there is any spatial correlation of
temperature, the coherence length scale must be smaller than $\sim
0.3h^{-1}$ Mpc to be consistent with the Keck data.

\end{abstract}


\altaffiltext{1}{{\sl Chandra} Fellow}

\section{Introduction}

In the last few years our understanding of small-scale structure at
high redshift has increased enormously through the study of the
Ly-$\alpha$ forest, the series of absorption features in the spectra
of distant QSOs. Based in large part upon insights gained from
numerical simulations (see, e.g., \citealp{cmo94,zan95,her96}) we now
know that the physics governing the Ly-$\alpha$ forest is relatively
simple.  At the relevant redshifts the gas making up the intergalactic
medium (IGM) is in photoionization equilibrium, which results in a
tight density--temperature relation for the absorbing material with
the neutral hydrogen density proportional to a power of the baryon
density.  Since pressure forces are sub-dominant, the neutral hydrogen
density closely traces the total matter density on the scales relevant
to the forest ($0.1-10\,h^{-1}$Mpc).  The structure in QSO absorption
thus traces, in a calculable way, slight fluctuations in the matter
density of the universe back along the line of sight to the QSO, with
most of the Ly-$\alpha$ forest arising from over-densities of a few
times the mean density (see, e.g., \citealp{cro98,cro02}).

While most attention has focused on the low-order statistics of the
Ly-$\alpha$ forest, the statistical properties of the flux are
significantly non-Gaussian. \citet{vie04} studied the bispectrum of a
large sample of QSO Ly-$\alpha$ flux spectrum. While their results
from observations and simulations and theoretical predictions
reasonably agree with each other at large scales, they found the
errors on the bispectrum are too large to discriminate between models
with different Ly-$\alpha$ forest distribution based on the bispectrum
method.

However, while in general probing high-$z$ physics with the bispectrum
is complicated, in some special cases patterns of higher order correlations
can be visible in the moments of the flux.  One particular pattern is
that due to the action of gravity in hierarchical structure formation.
The presence or absence of this pattern provides us with a powerful
probe of `extra' physical effects, beyond gravity, in the formation of
the Ly-$\alpha$ forest, as first emphasized by Zaldarriaga, Seljak \&
Hui (2001; hereafter ZSH01).  These authors noted that it was possible
to constrain fluctuations in the continuum by appealing to higher
order moments of the flux.  In this paper we wish to address another
physical effect: scatter in the density-temperature relation induced
for example by different thermodynamical histories for the gas,
fluctuations in the UV background radiation field or radiative
transfer effects.  We shall see that it is possible to put tight
constraints on the scatter in the $\rho-T$ relation by considering the
higher order moments of the spectra.

The outline of the paper is as follows. In \S\ref{sec:sim} we briefly
discuss the simulations involved in this work and how we extract
``mock'' Ly-$\alpha$ spectra.  The statistical method introduced by
ZSH01 is reviewed in \S\ref{sec:sta}.  Our main results are presented
in \S\ref{sec:sca}, and we discuss the application to high resolution
Keck spectra in section \S\ref{sec:app}. In \S\ref{sec:dis} we discuss
several related issues.

\section{Simulation} \label{sec:sim}

\begin{figure*}
\psfig{file=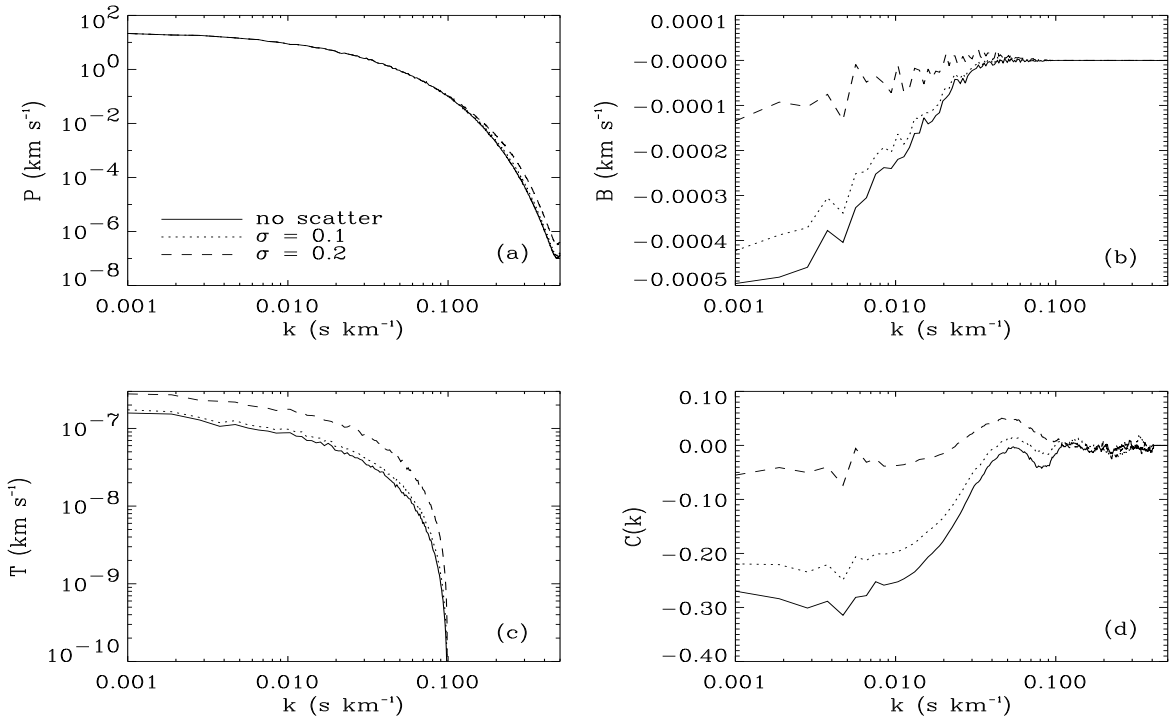,angle=90,width=0.85\textwidth,height=0.4\textheight}
\caption{Statistics of the Ly-$\alpha$ flux spectrum based on a square
filter $0.2\leq k\leq 0.3$.  Solid lines show the no scatter case,
dashed lines $\sigma=0.1$ and dotted lines $\sigma=0.2$.  Panel (a):
power spectrum; panel (b): bispectrum; panel (c): trispectrum; and
panel (d): cross-correlation coefficient $C(k)$. In panel (d) $C(k)$
is smoothed with a boxcar method for visual clarity.}
\label{fig:filter}
\end{figure*}

Since we are interested in the higher order correlations in the
Ly-$\alpha$ flux which are induced by gravitational processes, we
first need a model of gravitational clustering.  For this purpose we
use a particle mesh code to follow the evolution of dark matter
clustering, including only gravity.  Details of the PM simulation can
be found in \citet{mwp99}, \citet{whi99} and \citet{mwh03}.  Given the
particles' positions and velocities, the Ly-$\alpha$ forest spectrum
can be constructed based on the so-called ``fluctuating Gunn-Peterson
approximation'', in which Ly-$\alpha$ absorbing gas is assumed to trace
the dark matter distribution, and be in a state of photoionization
equilibrium with the UV background radiation (see, e.g.,
\citealp{cmo94,zan95,her96,hgn97}).  Under these assumptions the gas
temperature can be obtained through
\begin{equation}
  T = T_0\left(\frac{\rho}{\bar{\rho}}\right)^{\gamma-1}
\label{eq:rhoT}
\end{equation}
where $\rho$ is the matter density and $\bar{\rho}$ the cosmic mean.
The index $\gamma$ is expected to vary from near unity right after
reionization, to $\sim 1.5$ at low-$z$ \citep{hgn97}.  We will set
$T_0=2\times 10^4\,$K throughout, but in the next section we will
introduce scatter around the values given by Eq.~\ref{eq:rhoT}.

\begin{figure*}
\psfig{file=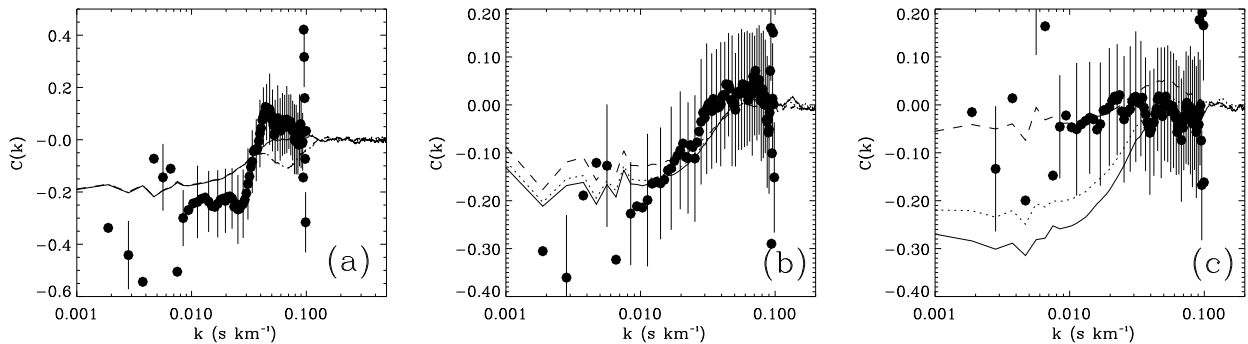,angle=0,width=0.9\textwidth,height=0.2\textheight}
\caption{Cross-correlation coefficient, $C(k)$, for different filtering
schemes.  Symbols are the same as those in Figure \protect\ref{fig:filter}.
Panel (a): $0.02 \leq k \leq 0.06$;
panel (b): $0.1 \leq k \leq 0.2$; and
panel (c): $0.2 \leq k \leq 0.3$. Keck data is plotted as the solid circle. Error bars are 1$\sigma$ and plotted every three points
for visual clarity.}
\label{fig:Ck}
\end{figure*}

The neutral hydrogen number density, based on our ionization
equilibrium assumption, is $n_{\rm HI} \propto
\left({\rho/\bar{\rho}}\right)^2\, T^{-0.7}$, with the constant of
proportionality dependent on the baryon density, redshift, and
hydrogen photoionization rate.  We shall fix it using the observed
value of the mean flux (see below).  With the gas density, temperature
and velocity data, the optical depth $\tau$, at a given velocity $u$ is
\begin{eqnarray}
  \tau &      =& \int n_{\rm HI}(x)\,\sigma(x)\,dx \nonumber \\
       &\propto& \int dx\ \left({\rho\over\bar{\rho}}\right)^2
       \,T^{-0.7}\,b^{-1}\exp\left[-{(u-u_0)^2\over b^2}\right]
\label{eq:tau}
\end{eqnarray}  
where $u_0$ is the sum of Hubble expansion and peculiar velocity at
position $x$, and $b = \sqrt{2k_BT/m_p}$ is the Doppler parameter.
Here $m_p$ is the proton mass, and $\sigma$ is the photoionization
cross section.  The flux at velocity $u$ is then $\exp(-\tau)$.

For definiteness we concentrate on the $z=3$ output of a $512^3$
particle and $1024^3$ mesh PM simulation in a box of side
$60\,h^{-1}$Mpc, which gives a simulation resolution of $\sim 6\,\rm
km\,s^{-1}$ at $z \sim 3$. The cosmology simulated was of the $\Lambda$CDM
family, with $\Omega_{\rm mat}=0.3$, $h=0.67$ and $\sigma_8=0.9$.  The
density and velocity fields were Gaussian smoothed by $100\,h^{-1}$kpc
to simulate the effects of gas pressure and $32^2$ spectra, each of
$1024$ pixels, were computed on a regular grid parallel to the box
axes.  The normalization constant was fixed to reproduced the mean
flux, $\bar{F}=0.684$, observed by \citet{mcd00} at $z=3$ (see also
Appendix B of \citealp{mwh04}).

\section{Statistics} \label{sec:sta}

To constrain scatter in the temperature density relationship on small
scales we adopt the high-order statistical method proposed by ZSH01.
We briefly describe this technique here and refer the reader to their
paper and \citet{mms03} for details.

We define the relative fluctuation of the flux, $\delta_F \equiv
F/\bar{F}-1$, and its Fourier transform
\begin{equation}
  F(k) = \int dx\,e^{ikx}\,\delta_F (x) \qquad .
\label{eq:Fk}
\end{equation}
We bandpass filter $\delta_F$ with a window function $W(k)$: $F_f(k) =
F(k)W(k)$, taking the window function $W(k)$ to be either a square
window,
\begin{equation}
  W_{k_1,k_2}(k) = \left\{
  \begin{array}{ll}  
1 & {\rm if}\,\,k_1 \leq k \leq k_2;\\ 0 & {\rm otherwise}
  \end{array} 
\right.
\label{eq:filter}
\end{equation} 
or a Gaussian window function.  Our results are not sensitive to this
choice, so in what follows we shall concentrate on the square window.

The filtered field $F_f(k)$ in Fourier space is then inverse
transformed back into real space to define
\begin{equation}
  \delta_H(x)= \int {dk\over 2\pi}\ e^{-ikx}\, F_f(k).
\label{deltaH}
\end{equation}
and this is squared to produce $h(x)\equiv \delta_H^2(x)$.  Following
the conventions in \citet{mms03}, we define $P$ as the power spectrum
of $\delta_F$, $T$ as the power spectrum of $h$, and $B$ as the
cross-spectrum between $\delta_F$ and $h$.  Since $h$ depends
quadratically on $\delta_F$, $B$ and $T$ can be related to the flux
bispectrum and trispectrum, respectively.  We also introduce the same
cross-correlation coefficient $C(k)$ to study the effect of large
scale density fluctuation on small scale power, defining $C\equiv B/\sqrt{PT}.$

\section{Scatter in $\rho-T$} \label{sec:sca}

We model the scatter in the temperature-density relation as log-normal
in $T$ at fixed $\rho$, with mean value determined from
Eq.~\ref{eq:rhoT}.  We choose the width, $\sigma$, to mimic the
results seen in hydrodynamic simulations.  Specifically we choose
$\sigma \equiv \delta\ln T=0.1$ and $\sigma=0.2$ to study the
effect. For comparison, by checking $\rho - T$ scatter plot, we find a
scatter of $\sigma = 0.1$(0.2) is consistent with results from low
(high) resolution hydrodynamic simulations in \citet{bma99},
respectively (see their Fig.~3). Figure~\ref{fig:filter}a shows the
power spectrum $P$ of our simulation under the assumptions of no
scatter, $\sigma=0.1$ or $\sigma=0.2$.  At $k \gtrsim
0.1\rm\,s\,km^{-1}$ (scales $\lesssim 10\,h^{-1}$Mpc), scatter in the
temperature-density relation adds power.

We show in Fig.~\ref{fig:filter} the power spectrum, bispectrum,
trispectrum and cross-correlation coefficient from the simulations
with a top-hat filter with modes between $0.2\le k\le
0.3\rm\,s\,km^{-1}$.  We chose that range to maximize the effect of
$\rho-T$ scatter on $C(k)$.  As pointed out by ZSH01, a positive
correlation between the large scale density fluctuations and the power
on small scales is induced by gravitational instability.  Since,
within the ``fluctuating Gunn-Peterson approximation'', the
fluctuations in the dark matter distribution are directly traced by
fluctuations in the Ly-$\alpha$ forest, the positive correlation
induced by gravitational instability is manifest in the Ly-$\alpha$
flux.  In the Ly-$\alpha$ forest the correlation is negative since the
flux decreases with increasing density.

We see that our model for scatter in the $\rho-T$ relation does not
change the large scale fluctuations of the Ly-$\alpha$ flux spectrum;
however, by locally varying the optical depth $\tau$ through
Eq.~\ref{eq:tau}, it effectively adds noise at small scales.
Figure~\ref{fig:filter}d shows $C(k)$ at low-$k$ becomes steadily less
negative as the scatter is increased. Note also that the bispectrum,
tri-spectrum and cross-correlation coefficient go quickly to zero for
$k \geq 0.1\rm\,s\,km^{-1}$.
This is because after filtering the highest mode in the squared field,
$h(k)$, is $k_2-k_1$, or $0.1\rm\,s\,km^{-1}$ in this case.

Finally, we have investigated the requirements on signal-to-noise and
resolution of spectra which would be suitable for measuring $C(k)$ on
the appropriate scales.  By adding Gaussian random pixel noise to our
mock spectra we found that a $S/N$ approaching 50 -- 100 is required
in order not to wash out the small scale signal.  We also notice that
the level of scatter is closely related to the simulation
resolution. The flux spectrum is obtained by averaging out
fluctuations in simulations cells over the thermal broadening
width. The smaller the cells are, the more cells will be averaging
out, and the smaller the fluctuation will be. In our simulation, the
typical temperature of the IGM is $\sim 2\times 10^4$ K, which
corresponds to a size of $\sim 18\,\rm km\,s^{-1}$. So the flux
spectrum in each pixel is roughly averaged over three neighboring
cells. To test this effect, we check simulations with two other
resolutions that are close to the ones in hydrodynamic simulations by
\citet{bma99}, which are two and four times higher than the one used
in the paper. We found that there are no significant differences among
these cases.

\section{An Application to Quasar Data} \label{sec:app}

In this section, we study the cross-correlation coefficient $C(k)$
with two, high resolution, high $S/N$ quasar spectra, Q~1422+231 at
$z = 3.62$ \citep{rau01} and HS~1946+7658 at $z=3.05$ \citep{kty97}.
These two quasars were observed with Keck High Resolution Echelle
Spectrometer (HIRES), with a signal-to-noise ratio ($S/N$) of $\sim 140$
and $\sim 50 - 100$, respectively.
We split the two spectra into a total of 13 separate pieces between
Ly-$\alpha$ and Ly-$\beta$ emission to avoid contamination from Ly-$\beta$
absorption.
Each piece has a comoving length of 60$h^{-1}$ Mpc to match our PM
simulation.
We calculate $C(k)$ among these 13 samples following the method described
in section \S\ref{sec:sta}, and compare these results against our PM
simulation.  We made no attempt to correct for metal lines.
In order to account for evolution in the mean flux, we adjusted each
spectral segment to $\langle F\rangle=0.684$ by scaling the implied
optical depths as in the simulations.

In Figure \ref{fig:Ck} we show $C(k)$ from both Keck data and our PM
simulation. To study the filter effect, we select three filters: (a):
$0.02 \leq k \leq 0.06\rm\,s\,km^{-1}$; (b): $0.1 \leq k \leq
0.2\rm\,s\,km^{-1}$; and (c): $0.2 \leq k \leq 0.3\rm\,s\,km^{-1}$.
With each filter we plot the no scatter case (solid lines), scattering
with $\sigma=0.1$ (dotted lines) and scattering with $\sigma=0.2$
(dashed lines). Keck data are labeled as solid circles, with error
bars computed from  the simulations (the sample size in the data is too
small to give constraints on variances).

Clearly, with large fluctuations, negative signals present in cases
with filter (a) \& (b), which agrees with simulation quite well. This
is consistent with what ZSH01 found in Q~1422+231 data, and confirms
the correlation between the large and small scale fluctuations of the
Ly-$\alpha$ flux spectrum induced by gravitational instability. Also
in case (a) \& (b) , there is almost no difference between no-scatter
and scatter because their power spectra (see Figure~\ref{fig:filter}a)
show no difference between $0.01$--$0.2\rm\,s\,km^{-1}$.
At smaller scales (larger $k$ modes) the scattering increases the power
spectrum and starts to differentiate between the scatter and no-scatter
cases.
We also find, with filter (c), the $C(k)$ from Keck data is essentially
zero, consistent with $20\%$ scatter in $\rho-T$ relation case.
Though other effects (such as random noise) may also drive $C(k)$ to zero,
increasing the $\rho-T$ scatter above 20\% drives $C(k)$ positive,
so we can conservatively set an upper limit of $20\%$ on the the scatter
in the $\rho-T$ relation.

In the above we adopted an independent random Gaussian-distributed $T$
for each pixel.  In reality, however, pixels tend to correlate with
neighbor pixels: pixels around a high $T$ pixel appear to also have a
high temperature and vice versa. Such a correlation among pixels tends
to produce a ``correlated'' noise from $\rho-T$ scatter and enhance
the signals in cross-correlation coefficient $C(k)$: the larger the
correlated pixel length scale, the stronger the $C(k)$.  So with
filter (c), we can ask: what is the largest length scale that pixels
can be correlated with each other, without showing a strong signal in
$C(k)$ that conflicts with the Keck data?  We can parameterize this by
adopting a correlation parameter $r$: $T_{i} = rT_{i-1} + (1-r)T_{\rm
new}$, while $T_i$ is the temperature of pixel $i$, $T_{\rm new}$ is
the temperature obtained from an independent random draw, and $r$ is a
number between 0 and 1: $r=0$ means pixels are independent of each
other, while $r=1$ mean the maximum correlation.  We found, for the
20\% scatter case, $r$ must be smaller than $\sim 0.3$ -- 0.5 not to
be ruled out by the Keck data.  This roughly corresponds to a distance
of $\sim 5$ pixels if we define two pixels correlated at the 1\% level
as ``correlated''.  This means, with a 20\% scatter, the correlated
scale length must be smaller than $\sim 0.3h^{-1}$ Mpc, or $\sim
30\rm\,km\,s^{-1}$.

The observational evidence for the gravitationally induced correlation,
and for small-scale decorrelation, currently rests on two high resolution
spectra.  It would obviously be of some interest to improve the analysis
presented here and to increase the sample of QSO spectra for which the
analysis has been performed.

\section{Discussion} \label{sec:dis}

ZSH01 showed that gravitational instability, and the assumption of the
``fluctuating Gunn-Peterson approximation'', predict a specific
correlation between large- and small-scale power in the Ly-$\alpha$
forest.  They also showed, based on high resolution Keck data, that
observations were consistent with the correlation predicted by
gravity.  Here we point out that this observation allows one to limit
the deviations {}from the ``fluctuating Gunn-Peterson approximation''
and in particular to constrain the fluctuations in the
density-temperature relation of the IGM.  Such fluctuations are
expected to arise from varying thermodynamical histories for the gas
and from fluctuations in the ultra-violet background radiation field
or radiative transfer effects.  We find that these fluctuations must
be $\lesssim 20\%$ in $T$ at fixed $\rho$ in order to be consistent
with high resolution Keck data.

Our method provides a sensitive way to probe for scatter in the
temperature density relation that is expected from the complex physics
of the IGM and seen in cosmological hydrodynamic simulations.
Persistence of gravitationally induced higher order correlations in
the spectra requires that the scatter be no more than twice as large
as already seen in hydrodynamic simulations of structure formation.
This limits the role of ``additional'' sources of fluctuations such as
an inhomogeneous UV background field or radiative transfer effects.
To further strengthen the case requires this correlation to be
computed from more high $S/N$, high resolution quasar spectra such as
could be obtained with Keck, Magellan, or similar telescopes.  We hope
that when more high quality spectra along different lines of sight
become available, this method will be a powerful tool to study any
``extra'' physical process, beyond gravity, in the IGM.

\smallskip
{\it Acknowledgments:} T.~Fang thanks Rupert Croft, Patrick Mcdonald,
Ravi Sheth and Matias Zaldarriaga for helpful conversations. We thank
M.~Rauch and W.~Salgent for providing Keck data of QSO 1422+231, and
A.~Meiksin for Keck data of HS 1946+7658. We also thank the referee
for comments on our manuscript. T.~Fang was supported by the NASA
through {\sl Chandra} Postdoctoral Fellowship Award Number PF3-40030
issued by the {\sl Chandra} X-ray Observatory Center, which is
operated by the Smithsonian Astrophysical Observatory for and on
behalf of the NASA under contract NAS8-39073. M.~White was supported
by the NSF and NASA. Parts of this work were done on the IBM-SP at the
National Energy Research Scientific Computing Center.

\end{document}